\documentstyle[12pt,aaspp4]{article}
\def\etal{{et al.}}
\def\asca{{\it ASCA}}

\def\gi{{\it Ginga}}

\def\xte{{\it RXTE}}

\def\ros{{\it ROSAT}}
\def\pcmsq{{$\rm cm^{-2}$}}
\def\chisq{{$\chi^{2}$}}

\def\delchi{{$\Delta \chi^{2}$}}
\def\Msun{\hbox{$\rm\thinspace M_{\odot}$}}

\begin{document}

\title {THE ORIGIN OF THE X-RAY AND ULTRAVIOLET EMISSION IN NGC 7469}

\author {K. Nandra\altaffilmark{1,2}, T. Le\altaffilmark{2,3},
I.M. George\altaffilmark{1,2}, R.A. Edelson\altaffilmark{4,5},
R.F. Mushotzky\altaffilmark{1}, B.M. Peterson\altaffilmark{6},
T.J. Turner\altaffilmark{1,7} }

\altaffiltext{1}{Laboratory for High Energy Astrophysics, Code 660, 
	NASA/Goddard Space Flight Center,
  	Greenbelt, MD 20771}
\altaffiltext{2}{Universities Space Research Association}
\altaffiltext{3}{Department of Physics and Astronomy, 
George Mason University, Fairfax, VA 22030}
\altaffiltext{4}{Department of Astronomy, University of California,
Los Angeles, CA 900024}
\altaffiltext{5}{X-ray Astronomy Group, Leicester University,
Leicester LE1 7RH, UK}
\altaffiltext{6}{Department of Astronomy, Ohio State University,
174 West 18th Avenue, Columbus, OH 43210}
\altaffiltext{7}{University of Maryland, Baltimore County, 
1000 Hilltop Circle, Baltimore, MD 21250}

\slugcomment{To be submitted for publication in
{\em The Astrophysical Journal}}

\begin{abstract}

We present a spectral analysis of a $\sim 30$~d, near-continuous
observation of the Seyfert 1 galaxy NGC 7469 with \xte.  Daily
integrations show strong spectral changes during the observation.  Our
main result is that we find the X-ray spectral index to be correlated
with the UV flux. Furthermore, the {\it broad-band} X-ray photon flux
is also correlated with the UV continuum. These correlations point
towards a model in which the X-rays originate via thermal
Comptonization of UV seed photons.  Furthermore, the UV is also
correlated with the extrapolation of the X-ray power law into the soft
X-ray/EUV region. Our data analysis therefore re-opens the possibility
the the UV photons and their variability arise from reprocessing, as
long as the primary source of heating is photo-electric absorption in
the reprocessor, rather than Compton downscattering, A coherent
picture of the X-ray/UV variability can therefore be constructed
whereby absorption and reprocessing of EUV/soft X-rays in a standard
accretion disk produce a variable seed photon distribution which are
in turn up-scattered into the X-ray band.  We also find a significant
correlation between the 2-10 keV flux and the 6.4 keV iron K$\alpha$ line
suggesting that at least some portion of the line originates within
$\sim 1$~lt day of the X-ray continuum source.  Neither the power law
photon index nor the Compton reflection component are correlated with
the 2-10 keV flux. The latter is not correlated with the iron
K$\alpha$ line flux either.  We do find an apparent correlation
between the X-ray spectral index and the strength of the Compton
reflection component.  In an Appendix we show, however, that this can
be produced by a combination of statistical and systematic errors. We
conclude the apparent variations in the Compton reflection component
may be an artifact of these effects.

\end{abstract}


\section{INTRODUCTION}
\label{Sec:Introduction}

Active galactic nuclei (AGN) emit most of their bolometric luminosity
in the ``big blue bump'' (BBB), which dominates the optical,
ultraviolet (UV) and Extreme Ultraviolet (EUV) emission, unless the
regions which produce these components are obscured. A further
ubiquitous property of AGN is hard X-ray emission, which is spectrally
distinct from the big blue bump and energetically significant. With
the exception of strong radio emission, which is only observed in a
small fraction of AGN, all other components of the AGN spectrum can
typically be accounted for by some kind of ``reprocessing'', by which
we mean passive absorption and re-radiation, without significant
additional energy input. Thus the BBB and X-ray continua can be
considered the ``primary'' emissions of AGN and determining their
origin holds the key to our understanding of the central engine of
active galaxies.

The evidence for black holes in AGN is now extremely strong
(e.g. Miyoshi et al. 1995; Tanaka et al. 1995) but the fueling and
emission mechanism are still not well known.  The probable presence of
angular momentum lead naturally to the hypothesis of an
optically-thick accretion disk around the black hole (e.g. Lynden-Bell
1969) which radiates the BBB (Shields 1979; Malkan \& Sargent 1982). A
disk geometry can also account for some optical line profiles
(e.g. Chen, Halpern \& Filippenko 1989; Eracleous \& Halpern 1994) and
explain the profiles of the broad, redshifted iron lines (e.g., Fabian
et al. 1995; Nandra et al. 1997).  Accretion disk models are not free
from problems, however. A significant difficulty is the presence of
extremely rapid, wavelength independent variations in flux
(e.g. Clavel et al. 1991; Peterson et al. 1991). Standard disks should show
time lags between optical and UV bands as instabilities propagate
through them, but these lags should be much longer than the observed
upper limits (e.g. Courvoisier \& Clavel 1991; Krolik et al. 1991;
Molendi, Maraschi \& Stella 1992).

X-ray illumination of the disk may offer a way of explaining these
rapid, wavelength-independent variations of the optical/UV
continuum. Guilbert \& Rees (1988) were first to suggest that dense,
optically thick material could reprocess and thermalize the X-ray
emission of AGN, re-emitting the radiation in the BBB. If that
material were the accretion disk, variations in the reprocessed flux
could occur on the order of the light-travel time, which is compatible
with the observations. An apparent time lag of $\sim 1$~d between the
short-wavelength UV and optical variations in NGC 7469 (Wanders et
al. 1997; Collier et al. 1998; Peterson et al. 1998; Kriss et al. 2000)
offers further support for this hypothesis. X-ray fluorescence and
Compton scattering in the disk then account for the iron lines and
hard tails observed in the X-ray spectra (Nandra \& Pounds 1994).

The illuminating hard X-ray emission is a separate component not
predicted by standard disk models and requires another emission
mechanism.  Currently-popular thinking is that the X-rays arise from a
``corona'' which Compton up-scatters softer ``seed'' photons
(e.g. Sunyaev \& Titarchuk 1980), presumably from the optical/UV
where there is an abundant photon field. Such models reproduce the
broadband characteristics of the X-ray spectrum (e.g. Haardt \&
Maraschi 1991, 1993), especially if the corona is patchy (Stern et
al. 1995). A patchy, flaring corona is also suggested by the
variability of the iron K$\alpha$ lines (Iwasawa et al. 1996, 1999;
Nandra et al. 1999). A high energy cutoff at $\sim 200$~keV observed
in some sources argues for a thermal distribution of upscattering
particles (Zdziarski et al. 1994; Gondek et al. 1996).

A reasonably coherent picture of the central regions of AGN can thus
be constructed where an accretion disk surrounds a supermassive black
hole, with active flaring regions above the disk. The BBB photons
produced from the thermal emission of the disk are Compton-up-scattered
in the flaring regions, producing the X-ray continuum. This continuum
then illuminates the disk, producing more (and rapidly variable)
optical/UV emission via reprocessing, the iron line via fluorescence,
and the reflection hump from Compton downscattering. 

In the simplest such picture, the UV and X-ray variations should be
strongly correlated as they are tied together by two mechanisms, the
upscattering of the UV photons producing the X-rays, and the
reprocessing of the X-rays into UV. Any time lags between the two
bands should be very small, being just the light travel time between
the regions, and their sense would indicate which of the two
mechanisms were the dominant one.  Some early observations did indeed
suggest a link between the UV and X-ray bands (Clavel et al. 1992;
Edelson et al. 1996) and between the UV and EUV (Marshall et al. 1997)
consistent with reprocessing. More recent observations have shown a
relationship between the EUV and X-rays, supporting Comptonization
(Chiang et al. 2000; Uttley et al. 2000).  Nandra et al. (1998;
hereafter N98), however, have presented a $\sim 1$~month,
near-continuous \xte\ monitoring campaign on NGC 7469, with
simultaneous UV data from IUE (Wanders et al. 1997). N98 found a poor
correlation between the 2-10 keV X-rays and short wavelength (1315
\AA) UV at zero lag, with strong X-ray variations on short timescales
but none in the UV. The maxima in the UV emission appeared to lag the
X-ray peaks by about 4d, but the minima were near-simultaneous. These
effects are difficult to explain in the context of the
relatively-simple and observationally attractive picture of the
central engine outlined above.

The NGC 7469 observations have apparently left both reprocessing and
Comptonization models at a dead end, at least for that source. Similar
behavior has been observed in NGC 3516, with a poor zero-lag
correlation between the 2-10 keV X-rays and the optical flux observed
both on short ($\sim$~day) and long ($\sim$ month) time scales
(Edelson et al. 2000; Maoz, Edelson \& Nandra 2000).  This invites
extrapolation to AGN as a class, and leaves contradictory and
conflicting evidence as to the existence of the accretion disk, the
importance of reprocessing and the origin of the X-rays. Here we
present a spectral analysis of the \xte\ data for NGC 7469 which, when
compared to the X-ray and UV continuum data, appears to resolve these
conflicts and contradictions. After describing the observations and
analysis technique briefly in Section 2, we discuss the \xte\ spectral
properties of NGC 7469 in Section 3. In Section 4 we discuss time
variability of the spectrum and the correlations between the X-ray
spectral properties and in Section 5 discussion the correlations of
those properties with the UV. In section 6 we discuss the results.

\section{OBSERVATIONS}

Many of the details of the \xte\ observations are described in N98.
The primary difference between their data reduction and ours is in the
employment of the new background model, known as the ``L7+activation''
model. This provides a reduction in the systematic errors, allowing
spectral analysis. We applied the standard selection criteria in the
``rex'' RXTE processing script. These excluded periods where the earth
elevation angle was less than 10 degree, where the offset between the
pointing position and the source was greater than 0.02 degree, a 30
minute period after passage through the SAA and times where anomalous
electron flares occurred. We restricted our analysis to the three PCU's
which were operating throughout the observation for reasons of
uniformity and simplicity.

Even with the new model, the background-subtraction is unreliable at
high energies for relatively weak sources such as NGC 7469. We have
therefore restricted our analysis to the 2-20 keV band, and consider
the possibility of systematic errors in subtraction in our analysis.
We calculated response matrices using PCARSP v2.37. We calculated
matrices for several different times of the observation, but found no
statistically significant difference between the derived spectra.  We
have therefore employed a single matrix for the entire observation,
calculated for a spectrum in the middle of the observational period.

The standard software estimates the error in the net (source minus
background) rate by combining the errors in the total rate with those
of the simulated background.  The background estimation software
calculates errors on the simulated background spectrum assuming
Poisson statistics appropriate for the period simulated. In practice
this is a considerable overestimate of the statistical error on the
background spectrum, which is estimated using a large amount of
data. Indeed, the statistical error in the background estimation
should typically be negligible compared to the error on the total
rate. We have therefore assumed zero error on the estimated
background, and calculate the error in the background-subtracted count
rate based only on the error in total rate. This is obviously not
strictly correct and will result in an underestimate of the
statistical errors which may become apparent in a very long
integration. More importantly, it also ignores any systematic error.
Later, we discuss in detail whether systematic errors in the
background subtraction are likely to affect our results.

In many cases below we have estimated X-ray power-law fluxes in bands
either broader than or far removed from the observed \xte\ (2-20 keV)
band. In such cases the error on the flux is dominated by the
uncertainty in the extrapolated spectral model.  In order to determine
the flux values and uncertainties we have therefore searched through
the parameter space defined by the 68 per cent confidence contours for
the spectral parameters, determining the minimum and maximum fluxes
consistent with the data at this confidence level.  From these we
adopt the mean of the maximum and minimum values as our flux estimate,
and the error to be half the difference between the maximum and minimum.

The IUE observations used here are described in Wanders et al. (1997)
and we have used the modified light curve derived by Kriss et
al. (2000) for all plots and correlations, which improves over the
original Wanders et al. light curve by taking account of low contrast
line features. In calculating UV fluxes for comparison with derived
X-ray spectral properties, we have taken an average of the (typically
3-5) IUE spectra taken during the X-ray integrations. For the error
bar on the UV points, we adopt the larger of the dispersion of the
individual IUE points, and the propagated Kriss et al. errors.

\section{Integrated X-ray spectrum}

The \gi\ data first showed the presence of an iron K$\alpha$ line and
reflection hump in NGC 7469 (Piro et al. 1990; Nandra \& Pounds 1994).
\asca\ confirmed the presence of the line, but a controversy remains
as to whether the line is broad or not, with Guainazzi et al. (1994)
finding no evidence for significant width and Nandra et al. (1997)
finding marginal evidence. This issue is largely irrelevant for \xte,
however, which has much poorer spectral response than the \asca\
SIS. We therefore fitted the spectrum of NGC 7469, integrated over the
entire 30 day period, with a model consisting of a power law, a
gaussian to represent the iron K$\alpha$ line and a reflection
component (pexrav; Magdziarz \& Zdziarski 1995). In the reflection
model, we assume no high-energy cutoff in the incident power law, and
that the reflection is from an optically thick slab inclined to our
line of sight with an inclination, $i$, corresponding to $\cos i =
0.95$, the lowest inclination allowed in the model.

\begin{figure}
\plotone{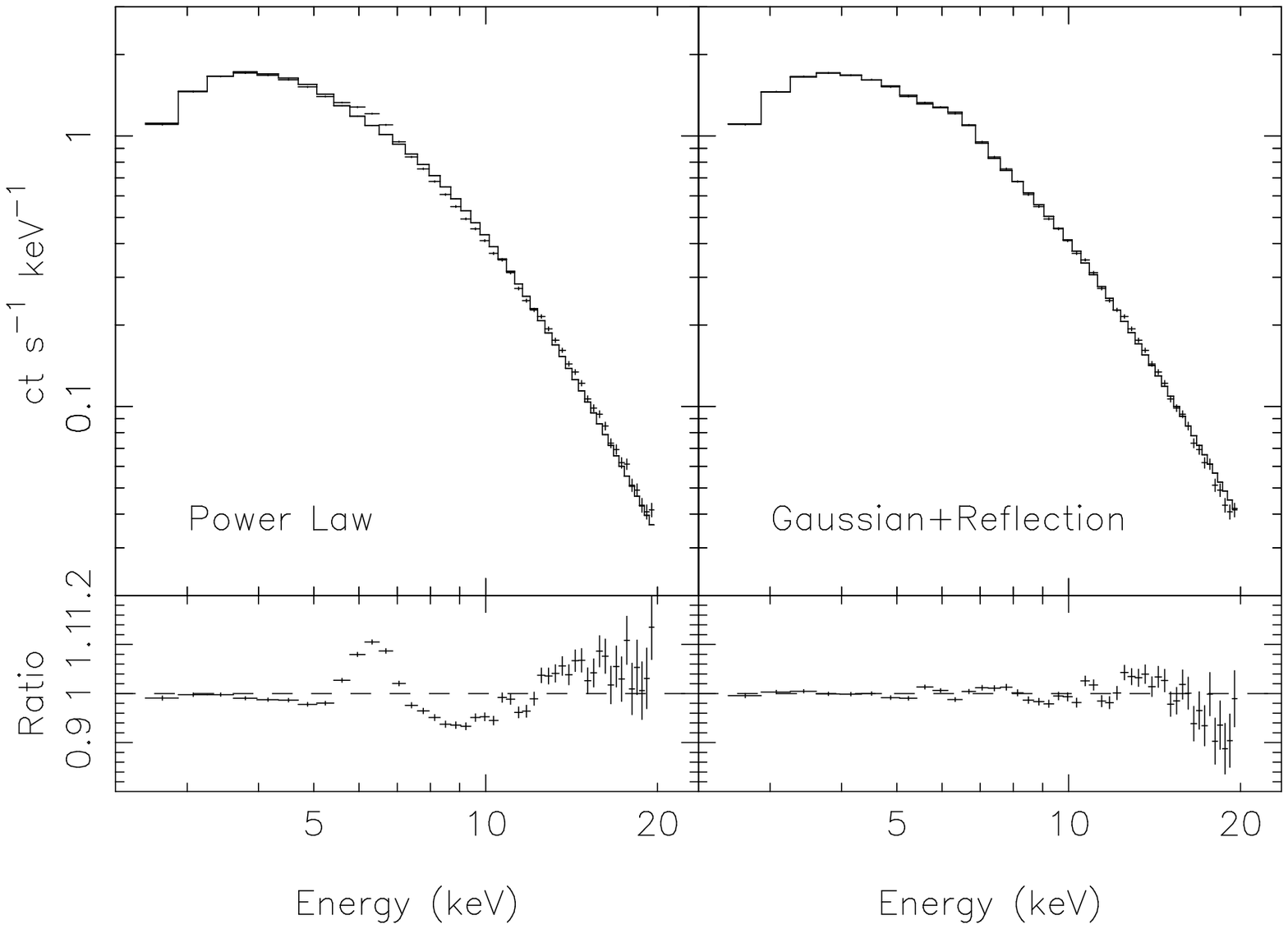}
\caption{
The left panels show a fit to the integrated spectrum with a power law
model. The upper panel shows the counts spectrum (crosses) and model
(line), after folding through the instrumental response. The lower
panel shows the ratio of this model to the data, clearly demonstrating
the presence of the iron K$\alpha$ line and reflection component.  The
fits is extremely poor, with a reduced \chisq\ of
$\chi_{\nu}^{2}$=47.6/44 d.o.f. The right panels show the the fit when
a gaussian line and reflection component are introduced into the
model. Though still not formally acceptable, perhaps due to systematic
effects, or the fact that the spectrum is variable (see
Fig.~\ref{fig:lc}), it is a vast improvement, with
$\chi^{2}_{\nu}=3.5$ for 41 d.o.f.
\label{fig:spec}}
\end{figure}

The spectral fits to a simple power law and this more complex model
are shown in Fig.~\ref{fig:spec}.  In these fits we fixed the
absorption column at the Galactic value, derived from 21 cm
measurements of $4.9 \times 10^{20}$~cm (Elvis, Lockman \& Wilkes
1989). The \xte\ data are insensitive to column densities $<$ few
$\times 10^{21}$~\pcmsq, for which there is no evidence from the
\asca\ or \ros\ spectra, which extend to lower energies (Brandt et
al. 1993; Guainazzi et al. 1994).  Although we did find a formal
improvement when the column was allowed to be free (\delchi=30), this
is due to a very small deviation ($\sim 2$~per cent) in the very
lowest energy bin.  We therefore consider the apparent absorption to
be due to small systematic effects in the calibration and/or
background subtraction.

In contrast the iron line and reflection hump represent much larger
deviations, both in percentage and \chisq\ senses. The inclusion of
the iron line reduces \chisq\ by 1600 compared to a power law, and the
Compton reflection further reduces it by \delchi=330. These are
obviously highly significant reductions in a statistical sense. The
deviations due to the iron line and reflection hump are of order 10
per cent (Fig. 1). The fit is not formally acceptable with
\chisq=144.7/41 d.o.f., which is rejected at very high confidence. The
residuals below about 15 keV are $<3$~per cent, however, and could be
due to the above-mentioned systematic errors. Another possible reason
for the poor fit is that, as we show below, NGC 7469 undergoes
substantial spectral variation during the observation. The combination
of several, dissimilar spectra would result in integrated data which
were difficult to fit. We have not therefore sought a more complex
model, and adopt the power law, gaussian and reflection fit as our
test model in investigating the spectral variability below. The
parameters from the integrated fit were a spectral index
$\Gamma=1.92\pm 0.02$, line energy $E_{K\alpha}=6.42 \pm 0.03$ keV and
line flux $F_{\rm K\alpha}=5.1\pm 0.3 \times 10^{-5}$~ph cm$^{-2}$
s$^{-1}$, corresponding to an equivalent width, EW = $150 \pm 10$~eV.
The line equivalent width is consistent with that seen by \asca\
(Guainazzi et al. 1994; Nandra et al. 1997).  The mean 2-10 keV flux
of $F_{\rm 2-10} = 3.28 \times 10^{-11}$~erg cm$^{-2}$ s$^{-1}$
corresponds to a 2-10 keV luminosity of $4.1 \times 10^{43}$~erg
s$^{-1}$.

The reflection normalization at 1 keV is $A_{ref}=5.1^{+1.0}_{-0.8}
\times 10^{-3}$ ph cm$^{-2}$ s$^{-1}$ keV$^{-1}$. The value of $A_{\rm
ref}$ corresponds to the normalization of the illuminating power law,
assuming 2$\pi$ coverage. We define the reflection fraction $R$ to be
the ratio of $A_{\rm ref}$ to the observed normalization of the power
law.  In other words, $R$ is the ratio of the strength of the
reflection component to that expected from a face-on slab subtending
2$\pi$ solid angle. For our data we find $R=0.48^{+0.09}_{-0.08}$.  On
the face if it, this suggests that the solid angle subtended by the
reflector is closer to $\pi$.  There are, however, numerous reasons
why the reflection fraction $R$ might be lower than unity. First, the
inclination might be higher than assumed (a face-on disk maximizes the
reflection component), with projection effects reducing the strength
of the reflection. If the assumed solid angle, spectral shape, and
power law illumination are correct, we can therefore derive an
inclination of $\cos i = 0.32^{+0.07}_{-0.06}$, corresponding to
$i=71{\arcdeg}\pm 4$.

Another possibility is that there is a cutoff in the primary spectrum,
which reduces the number of high energy photons able to be
down-scattered into the reflection hump. In this case we constrain the
cutoff energy $E_{\rm c} = 53^{+10}_{-7}$~keV, lower than is typical
for Seyfert galaxies (Madejski et al. 1995; Gondek et al. 1996), but
similar to, e.g., NGC 4151 (Maisack et al. 1993; Zdziarski, Johnson \&
Magdziarz 1996).  The introduction of an exponential cutoff into the
spectrum provides a significantly better fit (\delchi=42). Given the
uncertainties in the background modeling at the highest energies we
consider this to be inconclusive, particularly as the HEXTE spectrum
of this source appears to show a significant detection up to $\sim
70$~keV (R. Rothschild, priv. comm.). BeppoSAX also found no evidence
for such a cutoff, with a lower limit $E_{\rm c}>230$~keV (Matt 2000;
De Rosa et al. 2000). A final possibility is that the X-ray power law
is anisotropic (cf. Ghisellini et al. 1991), with the X-rays beamed
such that a factor $\sim R$ less flux travels towards the disk than
reaches the observer directly.

\section{Time-resolved spectral analysis}

The values derived from the integrated spectrum are largely consistent
with those found in the \gi\ observations of NGC 7469 (Piro et
al. 1990) and are rather typical of Seyfert 1 galaxies as a whole
(Nandra \& Pounds 1994). The real interest in our observation lies in
any variations. In the simplest reflection picture, we would expect
the line flux $F_{K\alpha}$ and the reflection flux $A_{\rm ref}$ to
be closely coupled to the continuum flux, and to each other. To
investigate this we have divided our observation into 30 segments of
approximately one day duration. These segments have exposure times
ranging from about 13 to 21 ks (Table~\ref{tab:var}) which are
sufficient to detect and constrain both the reflected flux and iron
line. We fitted each of these spectra with a model consisting of a
power law without high energy cutoff, a gaussian with energy fixed at
the value for neutral iron of 6.4~keV and reflection component with
free normalization $A_{\rm ref}$.  The results are shown in
Table~\ref{tab:var}. 

\subsection{Light Curves}

The light curves of $F_{\rm 2-10}$, $\Gamma$, $F_{K\alpha}$ and
$A_{ref}$ are shown in Figure~\ref{fig:lc}.  The bottom three panels
show the UV continuum light curve at 1315\AA, which we designate
$F_{\rm UV}$, and two other derived parameters which we discuss below.
The top panel shows that $F_{\rm 2-10}$ varies by $\sim 50$~per cent
during the observation (N98).  The photon index, $\Gamma$ is also
clearly variable, with \chisq=58.6/29 d.o.f against a constant
hypothesis (significant at $>99.9$~per cent confidence), but is not
simply related to the flux (see below). The iron line flux is
consistent with a constant (\chisq=13.4/29 d.o.f), but the best fit
values show trends similar to that of the continuum flux. The
reflection flux is again formally consistent with a constant
(\chisq=21.0/29 d.o.f), but appears to show trends similar to that of
$\Gamma$.

A more sensitive way of detecting variations in the parameters is to
use the F-test. We have repeated the fits to the daily spectra, but
this time fixing each of the parameters ($\Gamma$, $F_{K\alpha}$,
$A_{\rm ref}$) at their mean values (1.91, $5.1 \times 10^{-5}$, $5
\times 10^{-3}$). We then compute the F-statistic by comparing the
total \chisq\ of these fits to the \chisq\ when all parameters were
free. We derived F values of 5.38, 1.28 and 2.08 for 30 additional
parameters, implying highly significant changes in $\Gamma$ and
$A_{ref}$ ($>99$~ per cent confidence) but only marginal changes in
$F_{\rm K\alpha}$.  Nonetheless, we do believe there are variations in
$F_{\rm K\alpha}$ based on its correlation with the continuum. We now
explore the correlations between these X-ray parameters.

\subsection{X-ray/X-ray Correlations}

The results of linear (Pearson) and rank (Spearman) correlations
between the various X-ray parameters, assuming no time lags, are given
in Table~\ref{tab:cor}. The parameters are plotted against each other
in Fig.~\ref{fig:xcor}. Two positive correlations have apparent
significances of $>99$~per cent confidence, $F_{\rm 2-10}$ vs. $F_{\rm
K\alpha}$ and $\Gamma$ vs. $A_{ref}$. A marginal ($\sim 95$~per cent)
anti--correlation is seen between $A_{\rm ref}$ and $F_{\rm K\alpha}$.

\begin{figure}
\epsscale{0.82}
\plotone{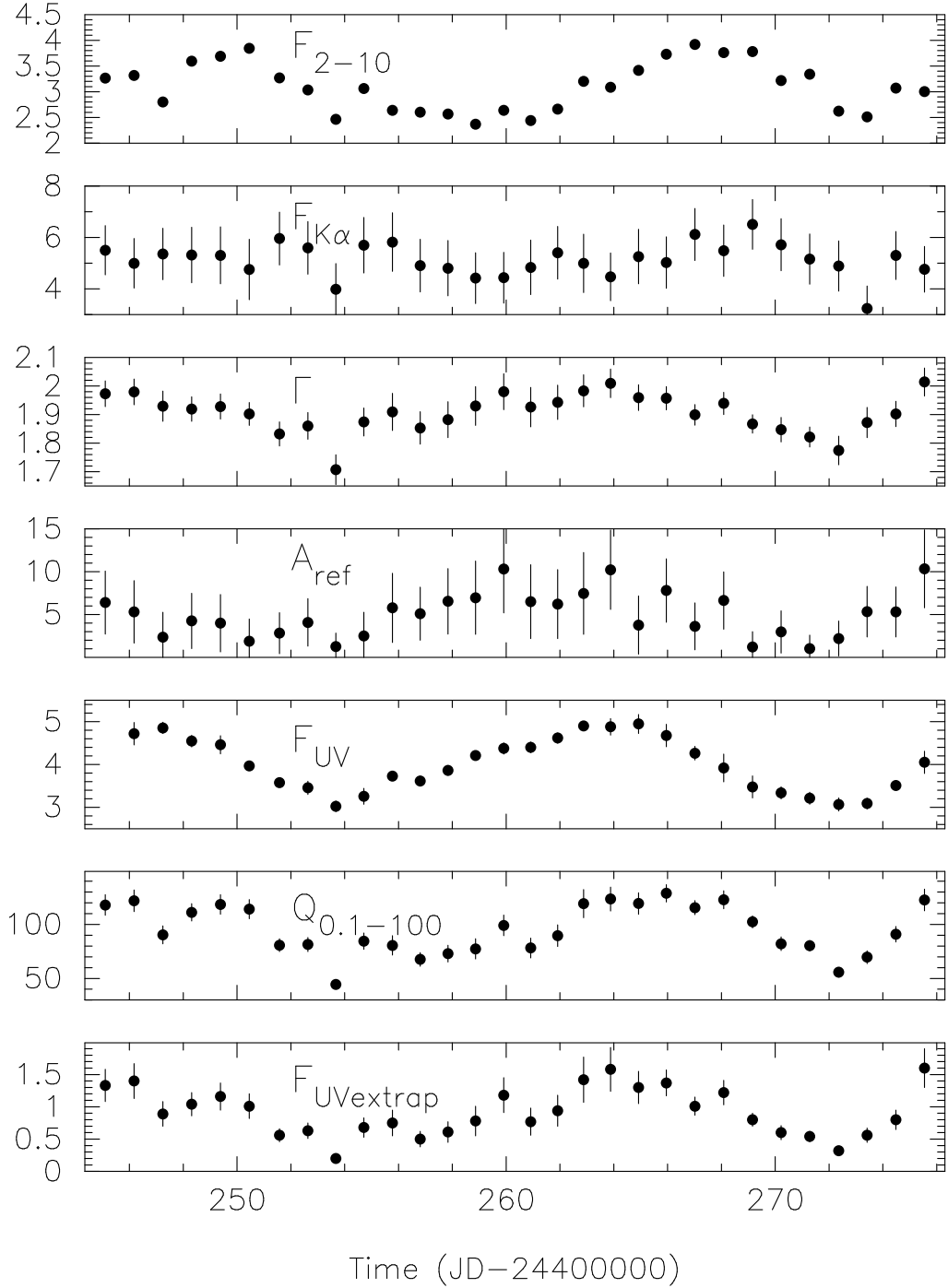}
\caption{Light curves of (descending), the 2-10 keV flux ($F_{\rm 2-10}$), 
the photon index ($\Gamma$), the iron line flux ($F_{\rm
K\alpha}$), the reflection flux ($A_{\rm ref}$), the UV continuum flux at
1315\AA\ ($F_{\rm UV}$), the broad-band X-ray photon flux ($Q_{\rm
0.1-100}$) and the X-ray power law flux extrapolated back to 1315\AA\
($F_{\rm UVextrap}$). The continuum index and reflection flux are
variable and not obviously related to the X-ray flux.  The iron line
is formally consistent with a constant but may be correlated with
$F_{\rm 2-10}$.  The light curves for $\Gamma$, $A_{\rm ref}$, $F_{\rm
UV}$, $Q_{\rm 0.1-100}$ and $F_{\rm UVextrap}$ all have similar shapes
and appear to be correlated with each other.  See
Figs.~\ref{fig:xcor}-\ref{fig:ucor}, Table~\ref{tab:cor} and the text
for further details.
\label{fig:lc}}
\end{figure}

\begin{figure}
\plotone{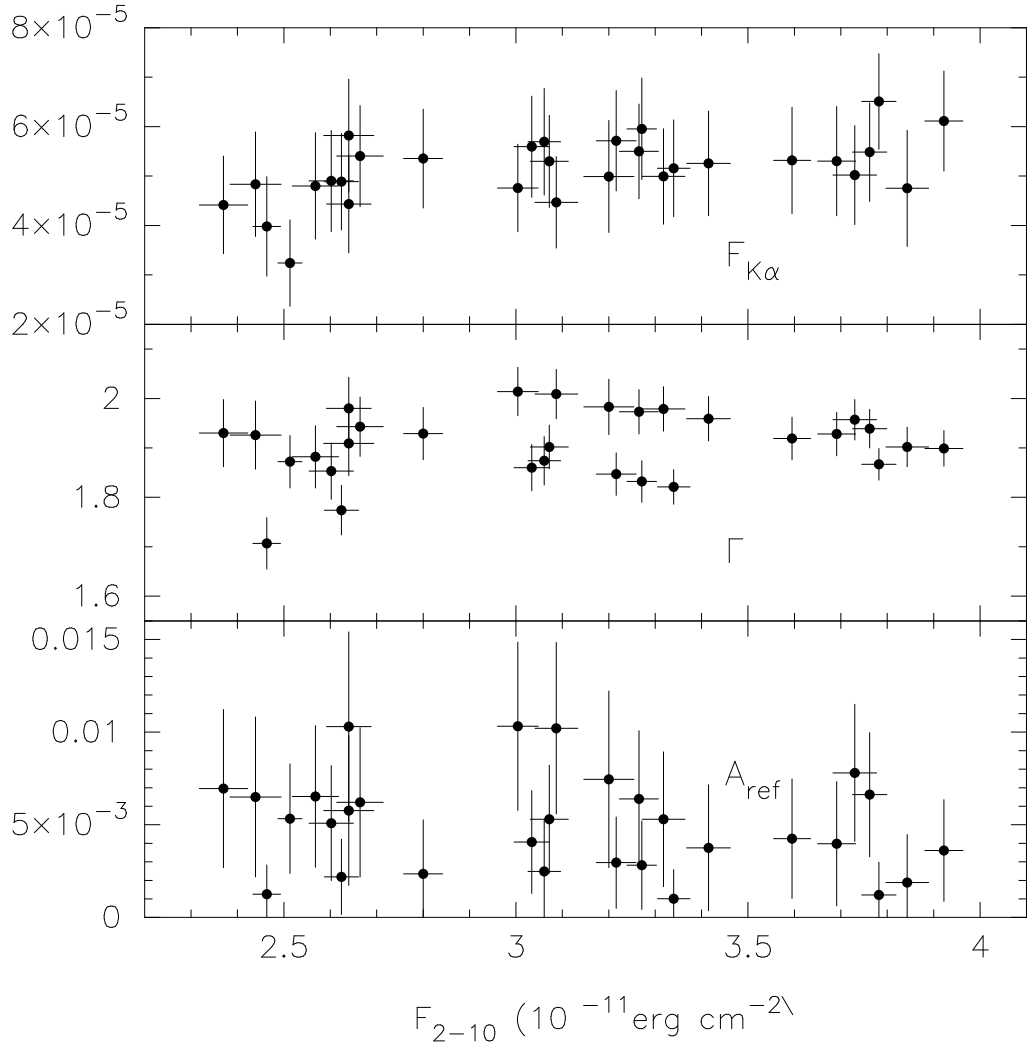}
\caption{Correlations between $F_{\rm 2-10}$ and a) line
normalization $F_{K\alpha}$, b) photon
index $\Gamma$ and c) reflection
normalization, $A_{\rm ref}$. The only parameter which shows a significant
correlation is the line normalization, at $\sim 99.8$~per cent
confidence (Table~\ref{tab:cor}).
\label{fig:xcor}}
\end{figure}

As discussed by Welsh (1999) and Maoz et al. (2000), cross-correlation
of light curves which have a ``red noise' character can result in
spuriously high values of the correlation coefficient due to the fact
that adjacent points in the light curve are highly correlated. This
could cause us to overestimate the significance of the correlations we
observe, due to the fact the the effective number of independent data
points in the correlation is less than the actual number of data
points.  We have tested this in the case of the $F_{\rm 2-10}$
vs. $F_{\rm K\alpha}$ correlation by simulating a number of
light curves with ``red noise'' power spectra (i.e. where the
variability power scales as $f^{-\alpha}$, where f is the
frequency). We adopt a value for the power law index of $\alpha=1.3$,
the best-fit value for a fit to the 2-10 keV power-density spectrum
presented by N98. As we have no a priori information regarding the
statistical characteristics of the line light curve, we have
correlated the simulated continuum light curves with the real light
curve of $F_{\rm K\alpha}$ (Fig.~\ref{fig:lc}). In 140 trials of 30
points, in no case did we find either a Pearson or Spearman
correlation coefficient as high as that observed with the real
data. The maximum values obtained were r=0.52 (Pearson) and r=0.51
(Spearman) compared to the r=0.54/0.52 found. We therefore conclude
that the correlation between $F_{\rm 2-10}$ and $F_{\rm K\alpha}$ is
unlikely to arise by chance coincidence.

The correlation we observe between the line and continuum fluxes
indicates that the line emission comes from very close to the central
regions, consistent with current thinking about the line origin
(e.g. Tanaka et al. 1995; Nandra et al. 1997). Ideally, we would use
such data to perform ``reverberation mapping'' of the inner accretion
disk, similar to that which has been performed in the optical/UV
(e.g. Blandford \& McKee, 1982; Peterson 1993).  With our current
sampling and data quality this is not possible, but as these are
arguably the best data obtained thus far for such a purpose, we have
calculated the interpolation cross correlation function (Gaskell \&
Peterson 1987; White \& Peterson 1994) between the line and continuum
(Fig~\ref{fig:ccf}). The data are not of sufficient quality to place
limits on the lag using the methods of, e.g. Peterson et
al. (1998). Nonetheless, the fact that a significant correlation is
observed is indicative that the line arises from within a few light
days of the continuum source.

It is apparent from Fig.~\ref{fig:lc} that both $\Gamma$ and $A_{\rm
ref}$ are uncorrelated with the 2-10 keV continuum flux, confirmed by
Fig.~\ref{fig:xcor}.  Earlier observations, mostly with poorer data
quality and sampling, had shown a tendency for a correlation between
$\Gamma$ and the X-ray flux (e.g. Turner 1987; Matsuoka et al. 1990;
Leighly et al. 1996).  As we shall show below, our data show that the
situation may be more complex than a simple correlation between
$\Gamma$ and X-ray flux. Indeed the $\Gamma$ light curve of NGC 7469
is reminiscent of the {\it ultraviolet} emission of NGC 7469 which we
also plot in the Fig.~\ref{fig:lc}, along with two other quantities
derived from the X-ray spectra. We explore these further below.  

The lack of correlation between $A_{\rm ref}$ and the $F_{\rm 2-10}$,
or $A_{\rm ref}$ and $F_{\rm K \alpha}$ is difficult to explain in the
standard picture in which both arise from the same, small region
(e.g. George \& Fabian 1991; Matt, Perola \& Piro 1991). We have
reason, however, to believe that the derived values of $A_{\rm ref}$
may not be truly representative of the strength of the reflection
component. 

A high correlation coefficient is obtained for $\Gamma$ vs. $A_{\rm
ref}$ (Table 1; see also Zdziarski, Lubinski \& Smith 1999), but great
caution must be exercised when attempting to correlate spectral
parameters which were derived from a single fit with a low spectral
resolution detector.  This is because the parameters are often
correlated in the fit, and thus a particular value of one might lead
statistically and systematically to a certain value of another (see
also Matt 2000). We discuss this further in Appendix A, where we have
performed simulations which show that a combination of statistical and
systematic errors can account for the observed correlation. We
therefore assign no statistical confidence to the $\Gamma$/$A_{\rm
ref}$ correlation and make no explicit interpretation of it. We further
assign no significance to the lack of correlations between $A_{\rm ref}$
and $F_{\rm 2-10}$ or $F_{\rm K\alpha}$.

\section{X-ray/UV correlations}

Fig.~\ref{fig:ucor} shows a number of X-ray spectral parameters and
fluxes plotted against the UV continuum flux (Wanders et al. 1997;
Kriss et al. 2000). Correlation coefficients are shown in
Table~\ref{tab:cor}.  There is a clear correlation between $F_{\rm
UV}$ and the X-ray spectral index $\Gamma$. We also find a marginal
correlation between the UV continuum and reflection component, $A_{\rm
ref}$, but this is most likely to be an artifact of the (real)
correlation between $\Gamma$ and $F_{\rm UV}$, combined with the
(possibly spurious) correlation between $\Gamma$ and $A_{\rm ref}$
(see Appendix A). We have again performed simulations to test whether
the $F_{\rm UV}$--$\Gamma$ correlation is due to chance coincidence when
dealing with red noise light curves.  This time we adopt an index
$\alpha=1.9$ for the simulated light curves (the best-fit value for
the $F_{\rm UV}$ PDS of N98) and correlated them with the real
$\Gamma$ light curve. Once again, with 140 trials, we failed to find a
correlation coefficient as significant as that which we obtained. Here
we found maximum values of 0.69/0.69 (linear/rank) compared to
0.81/0.89 observed.
 
We now consider some possible interpretations of the $\Gamma$/$F_{\rm
UV}$ correlation. The first is that the UV flux is simply an
extrapolation of the X-ray power law to 1315\AA. We have tested this
by performing such an extrapolation, and the derived values $F_{\rm
UVextrap}$ are plotted in Fig.~\ref{fig:lc} and against $F_{\rm UV}$
in Fig.~\ref{fig:ucor}.  We indeed see a correlation between the two,
but $F_{\rm UVextrap}$ is a factor $\sim 3$ less than the observed
flux. One explanation for this is that the extrapolated (and variable)
X-ray component lies on top of a stronger uncorrelated component
(perhaps the disk flux), in this case having a flux of $\sim 3 \times
10^{-15}$~erg cm$^{-2}$ \AA$^{-1}$.  The addition of such a component
to the extrapolated X-ray flux almost reproduces the observed UV light
curve, but the amplitude of variation of the real light curve is greater
than the extrapolated one. Furthermore, in the general case (though
not necessarily in NGC 7469), the UV shows a flatter spectral slope
than the X-rays.  We conclude that a simple extrapolation of the X-ray
spectrum is unlikely to be the origin of the UV flux.

Nonetheless we note with great interest that if the soft X-ray/EUV
flux of NGC 7469 is a simple extrapolation of the harder X-ray power
law we measure with \xte, the soft X-rays and UV would be highly
correlated. Such a correlation could revive reprocessing models for
the UV, as long as the reprocessing was primarily of soft X-rays.  The
best way of establishing the connection between the soft X-ray and UV
is of course to correlate them directly.  Our \xte\ data have a lower
threshold of $\sim 2$~keV and are therefore not ideal for this
purpose. We have, however, attempted this in a comparative way by
correlating the \xte\ 2-4 keV and 7-12 keV fluxes (which have similar
signal-to-nose ratios) with $F_{\rm UV}$. We did find a higher
correlation coefficient ($r=0.34$ for a Pearson correlation) for the
soft X-rays than for the hard ($r=0.07$). The soft X-ray correlation
is not significant, however, making it difficult to draw firm
conclusions from this analysis.  Our conclusion of a soft X-ray/UV
correlation therefore relies on the model-dependent assumption that
the hard X-ray power law extrapolates to lower energies.  While we
note that, if there are additional soft X-ray emission components, this
may not be valid the \asca\ and \ros\ data both indicate that a
single component dominates the emission over the whole X-ray band. We
therefore offer below an interpretation of the correlation between
the extrapolated soft X-ray flux and $F_{\rm UV}$.

Another reason why the UV flux and X-ray spectral index could be
related is if the X-rays arise from Comptonization of the UV photons
in a corona. When the UV flux increases, this should cool the corona
resulting in a softer Comptonized spectrum. This was one of the
possibilities that the UV/X-ray variability campaign was designed to
test, and in such a scenario we expect a correlation between the X-ray
and UV fluxes. In the presence of spectral variability, however, as we
observe here, it is not clear whether a given narrow-band (e.g. 2-10
keV) flux should show the correlation. In the Comptonization process
energy is added to the seed photons by the corona, which eventually
emerge at some X-ray energy. For a cool corona, more photons would
emerge at low energies, while a hotter one would have relatively more
hard X-ray photons. Regardless of the properties of the corona,
however, we should expect each UV photon up-scattered to result in one
emerging X-ray photon. We would therefore expect a correlation between
the UV flux and the broad band X-ray photon flux. We plot the 0.1-100
keV photon flux, $Q_{\rm 0.1-100}$ in Figs~\ref{fig:lc} and
\ref{fig:ucor} also. It can be seen that this is also correlated with
the UV flux (and also with $\Gamma$ and $F_{\rm UVextrap}$), and this
again supports the Comptonization model.

Confirming the results of N98, we find no significant correlation
between $F_{\rm 2-10}$ and $F_{\rm UV}$. We neither find any
relationship between $F_{\rm UV}$ and $F_{\rm K\alpha}$.

\section{DISCUSSION}

Spectral analysis of the \xte\ data during the 30 day simultaneous
campaign with IUE has revealed further important information regarding
the X-ray and UV emission in AGN.  A significant correlation is
observed between the 2-10 keV X-ray flux and the flux of the iron
K$\alpha$ line.  The other significant correlation observed within the
X-rays is between $\Gamma$ and the strength of the reflection
component, similar to that seen by Zdziarski et al. (1999) when
comparing the properties of a heterogeneous sample of AGN.  In our
\xte\ data, such a correlation can be produced spuriously, however, by
a combination of statistical and systematic effects, and we assign no
confidence to it or the (most likely secondary) correlation between
$A_{\rm ref}$ and the UV flux (Appendix A). We do find a significant
correlation between the X-ray spectral index and the UV flux of NGC
7469. We also find that the broad-band X-ray photon flux is well
correlated with the UV, as is the X-ray power law flux extrapolated
back into the EUV/soft X-ray band.  We now discuss these results and
what they tell us about the nature of the emission mechanisms and
reprocessing region in NGC 7469.

\subsection{Origin of the X-ray continuum}

Our data strongly support the idea that the X-rays in NGC 7469 are
produced by Compton upscattering of UV photons.  We account for the
changing spectral slope by hypothesizing that an increase in the UV
seed photons cools the Comptonizing corona and producing a soft X-ray
power law slope (e.g. Haardt, Maraschi \& Ghisselini 1994; Zdziarski
et al. 1999). We also observed a correlation between the broad-band
X-ray photon flux and and the observed UV flux. This is also expected
in Comptonization models, where an increase in the UV photon field
should provide a corresponding increase in the number of X-ray photons
which eventually emerge from the corona.  The behavior of NGC 7469 in
this respect is similar to that observed in the state changes of
galactic black-hole candidates (GBHC; e.g. Ebisawa, Titarchuk \&
Chakrabati 1996; Gierlinski et al. 1999), which show a steeper hard
X-ray slope in the ``high'' state, where the soft X-ray flux is
dominant.

\subsection{Origin of the UV continuum}

As we have mentioned above, a possible connection could exist between
the X-ray and UV fluxes if the latter is simply an extrapolation of
the X-ray spectrum. This might be the case if a broadband continuum
were generated by synchrotron or synchrotron-self-Compton emission, as
is thought to be the case in blazars.  Such a model has an immediate
attraction, as it can explain rapid, wavelength-independent variations
in the UV without the problems associated with an accretion disk.  In
its simplest form, we can rule out this model as the extrapolated
X-ray spectrum under-predicts the UV flux by a large factor.  A
further refinement of this idea is to introduce an additional,
near-constant component to the UV. This could arise, for example, from
the accretion disk and when added to the extrapolated flux could
account for the emission.  We again find this unlikely as then the
amplitude of variability of the extrapolated flux is less than that
observed. A further difficulty for this model is that the 2-10 keV
X-rays show very rapid variations which are not observed in the UV
(N98; see also Welsh et al. 1998). This is hard to account for if they
both represent the same component.

Indeed, our data strongly suggest that the UV emission - or at least
the variable part of the UV - arises from reprocessing of soft X-ray
and/or EUV photons. N98 had concluded that reprocessing was unlikely,
based on the fact that $F_{\rm 2-10}$ and $F_{\rm UV}$ were poorly
correlated at zero lag. Reprocessing models do indeed predict a
correlation between the X-rays and UV but our spectral data have shown
that the lack of a correlation in the case of NGC 7469 may have been
do to the fact that we were observing a variable X-ray spectrum over a
limited bandpass. In the Compton reflection/reprocessing scenario, two
mechanisms can heat the reprocessing medium: absorption of soft
X-rays, or Compton scattering of hard X-rays. The assumption of N98
was that the 2-10 keV X-ray flux was a good measure of these
mechanisms. The spectral data have shown that this is not necessarily
the case and, in particular, if the heating of the reprocessor is
dominated by absorption of soft X-ray and EUV photons, we find our
data to be entirely consistent with the reprocessing scenario. Soft
X-ray absorption is a more efficient heating process than Compton
scattering, as the cross sections are large below $\sim 10$ keV, and
the photons deposit all of their energy when absorbed, compared to
only a fraction $E / m_{\rm e} c^{2}$ for scattering, where $E$
is the photon energy.

Scattering can dominate the heating, however, in cases either where
the reprocessor is highly ionized - reducing the absorption
cross-section for soft X-rays - or if the intrinsic spectrum is very
flat such that the hard X-rays are energetically dominant.  In either
scenario, even with a variable spectrum, one would expect the
medium/hard X-rays ($>10$~keV) to dominate the reprocessing. In NGC
7469 we see a range of spectral indices $\Gamma\sim 1.7-2.0$. Assuming
an X-ray spectrum extending between 0.1 and 100 keV and a near-neutral
reprocessor, we would expect soft X-ray absorption to dominate in the
soft state (i.e. when the UV flux is high) but to have more equal
contributions in the harder state (i.e. where the UV flux is weak).
For the soft state, which has $\Gamma=2.0$, the 0.1-10 keV flux (which
can be absorbed) is a factor $\sim 2$ greater than 10-100 keV flux
(which is more likely to scatter). In the harder state, with
$\Gamma=1.7$, they are comparable.  This accounts for the fact that
N98 observed the minima in $F_{\rm 2-10}$ and $F_{UV}$ simultaneously,
as at those times $F_{\rm 2-10}$ is a good representation of the
heating flux. At high UV fluxes (and therefore steep X-ray slopes),
however, soft X-ray heating dominates, and $F_{\rm 2-10}$ is not a
good representation of the heating.  This accounts for the lack of
simultaneity between the maxima in $F_{\rm 2-10}$ and $F_{\rm UV}$
noted by N98.

N98 estimated the luminosity available for reprocessing, showing that
it was just sufficient to account for the UV flux at the shortest
(observed) wavelength of 1315\AA. The luminosity was not, however,
sufficient to account for the whole of the optical/UV continuum. In
practice, the reprocessing only needs to account (and in some
geometries {\it should} only account) for the variable part of the
optical/UV emissions. To estimate the luminosity of the variable
component, we have created a ``difference'' spectrum of the variations
by subtracting the approximate maximum and minimum fluxes of NGC 7469
in the optical/UV (Collier et al. 1998; Kriss et al. 2000). The
observational data cover the wavelength range 1315\AA-6850\AA. We then
fitted this difference spectrum with a power law, and integrated it
from 912\AA - 10,000\AA, obtaining a peak-to-peak variable flux of
$4.3 \times 10^{-11}$~erg cm$^{-2}$ s$^{-1}$. The difference between
the estimated X-ray fluxes in the 0.1-10 keV band at these times was
$4.1 \times 10^{-11}$~erg cm$^{-2}$ s$^{-1}$ (we crudely assume the
X--rays above 10 keV are scattered, rather than absorbed). These are
obviously comparable, although a fair comparison would require
knowledge of the geometry and ionization of the reprocessor. We
conclude, however, that soft X-ray reprocessing is an energetically
plausible source of the variable UV emission.

\subsection{Nature of the reprocessing regions}

The observed correlation between $F_{\rm 2-10}$ and $F_{K\alpha}$ does
not suffer from the complexities mentioned above, as the 2-10 keV flux
should closely follow the $\sim 7-10$ keV flux which excites the
emission line. Although we established a significant correlation at
zero lag, we were unable to set limits on the lag due to very low
signal-to-noise ratio in the line. The observed correlation implies
that at least some part of the line comes from very close to the
central source. This supports current thinking that the line arises
due to X-ray illumination of the inner accretion disk, based on the
line profiles (e.g. Tanaka et al. 1995; Nandra et al. 1997).  There is
no high signal-to-noise ratio profile of the iron K$\alpha$ line in
NGC 7469, so there remains controversy as to whether it exhibits the
characteristic, broad profile of an accretion disk. Our variability
data are certainly consistent with there being a constant component to
the line from material further away from the nucleus, such as the
optical BLR (e.g. Holt et al. 1980) or the molecular torus envisaged
in Seyfert 1/2 unification schemes (e.g. Ghisellini, Haardt \& Matt
1993; Krolik, Madau \& Zycki 1993).

Because of the potential systematic effects mentioned in appendix A, we
can draw no conclusion about the origin or location of the reflection 
component. The two simplest cases have the reflection arising in
the accretion disk, in which case it should be well correlated with
the hard X-ray flux and the iron line, or the molecular torus,
when it would remain roughly constant in strength and shape during
the period of our observation. Neither is suggested by the data if
taken at face value, but we await future data with better sensitivity
and smaller systematic errors before drawing any conclusion. 

If the variable UV emission does indeed arise from reprocessing of
soft X-ray and EUV photons, as seems likely, we can make some
inferences regarding the reprocessing region. The fact that no rapid
variability is observed in the UV, while the X-rays show
large-amplitude flares on short time scales (N98) indicates that the
UV comes from a region more extended than the X-rays. Light-travel
effects could then smear the fast variations. Berkley, Kazanas \&
Ozik (2000) have attempted to model the X-ray/UV variations of NGC
7469 taking into account the smearing by an accretion disk. They were
unable to smooth out the fast X-ray variations in such a way as to
reproduce the UV light curve. This may also provide difficulties for
the soft X-ray-UV reprocessing model proposed here, although the
Berkley et al.  calculations would have to be repeated in the light of
the spectral variations we observe here. It should also be noted that
we have not actually observed rapid soft X-ray/EUV variations, merely
inferred them from the presence of harder X-ray variability. If the
rapid flares in the X-ray light curve seen by N98 had hard spectra
(e.g. if they were absorbed), they might not have a measurable
effect on the UV light curve.

It is apparent from the spectral energy distribution of N98 that the
overall energetics of NGC 7469 cannot be dominated by reprocessed
X-rays. In particular, the starburst and other galactic emission 
make a large contribution in the IR and optical
(Genzel et al. 1995). It also seems likely
that there is a relatively--steady contribution to the UV emission, perhaps 
intrinsic thermal emission from the accretion disk. These should
also act as a seed source for the X-rays and one could envisage
either UV emission mechanism being dominant at any given time
in a single source, or when comparing different sources. It appears
that in NGC 7469, at least at the epoch observed, the two mechanisms
made rather similar contributions to the short-wavelength UV flux.

\subsection{Origin of the variability} 

A significant problem for standard accretion disk models has been
their inability to reproduce the observed rapid variations in the
optical/UV, which show at most small time lags at different
wavelengths (Clavel et al. 1991). This is not a problem for
reprocessing models, as variations can occur as fast as the light
travel time (e.g. Clavel et al. 1992). Our data show that reprocessing
in a standard accretion disk can account for the observed UV
variability data in NGC 7469, as long as the dominant heating process
is absorption of soft X-rays. Thus variations in the soft X-ray flux
could drive the UV variations.

As we now also have strong evidence, however, that the X-rays are
produced by Compton upscattering of UV photons, the inverse could hold
- in other words that UV variations drive those in the X-rays.
Establishing which requires measurement of a time lag between the
variations, which we have been unable to constrain. In practice,
however, it seems likely that a ``feedback'' mechanism operates, with
each band driving the other to some degree (Haardt \& Maraschi 1993).

Such a ``feedback'' mechanism cannot obviously account for the rapid
variations seen in $F_{\rm 2-10}$. Also, although we believe we have
identified the radiation process which produces the X-rays
(i.e. thermal Comptonization), the mechanism by which the X-ray corona
is heated remains mysterious. Presumably that mechanism could induce
intrinsic (and potentially rapid) X-ray variability, as noted by N98.
There are very few specific models for heating the X-ray corona.  Hot
accretion disk/ADAF models (Shapiro, Lightman \& Eardley 1976; Narayan
\& Yi 1994) heat the accreting gas to X-ray temperatures naturally,
but would also require a standard disk to co-exist with the hotter
flow to provide the optical/UV photon field and the reprocessing
medium (e.g. Lasota et al. 1996; Gammie, Narayan \& Blandford 1999). A
serious problem for ADAFs is that the variability timescale is
expected to be longer than that observed.  For reasonable black hole
masses of $10^{7-8}$~\Msun, the sound-crossing time is of order
10-100ks (Ptak et al. 1998). N98 observed large-amplitude changes on
time scales at the lower end of this range, making an ADAF unlikely.
It has also been suggested that flaring regions heated by magnetic
reconnection might be the source of the X-ray scattering plasma
(e.g. Nayakshin \& Melia, 1997; Poutanen \& Fabian 1999). Such regions
might be highly chaotic and unstable, and could produce rapid
intrinsic variability. We hope our new results will encourage a more
quantitative comparison of these models with the available data.

\subsection{Comparison with other campaigns}

There has been some discussion (e.g. N98; Edelson et al. 2000) about
the apparent discrepancies between multi-waveband variability
campaigns for Seyfert galaxies. Some have found significant
correlations (e.g. Clavel et al. 1992; Edelson et al. 1996), while
others have apparently shown weak, complex or null relations
(e.g. Done et al. 1990; Marshall et al. 1997; N98; Maoz et al. 2000;
Edelson et al. 2000). Our data have shown that, before a complete
picture of the multi-waveband variability can emerge, one must
consider the X-ray spectral properties. Such considerations could
easily account for the apparently discrepant behavior between objects,
and the apparent complexities observed in individual sources. In the
presence of X-ray spectral variability, the source of reprocessing
photons can change between the hard and soft X-ray bands.  If the
reprocessing picture outlined above is correct, the interband
relationships could vary in complexity, depending sensitively on the
exact spectral shape in the X--rays and the ionization state of the
reprocessor, both of which can be variable. Roughly speaking, we
expect objects with flat X-ray spectra ($\Gamma \ll 2.0$) to show a good
correlation between the hard X-ray and UV variability, as Compton
downscattering dominates the heating. If the X-ray spectrum is
steeper, photoelectric heating will dominate, and we then predict that
the soft X-rays should correlate with the UV. Of course, if there is a
high energy cutoff in the X-ray spectrum this would reduce the
importance of Compton heating, as substantial ionization of the
reprocessor would to photoelectric absorption.  Presumably
spectral changes due to the radiation processes could occur in the UV
and optical too, making it all the more dangerous to make inferences
from narrow-band observations.

To take some specific examples, we note that in NGC 4151 (Edelson et
al. 1996), the best-sampled multi-wavelength campaign before the
current dataset was obtained, a correlation was observed between the
{\it soft} X-rays and the UV, just as we infer here. NGC 4151 has a
relatively flat X-ray spectrum (Yaqoob \& Warwick 1991), but also
exhibits a cutoff in the primary spectrum at a relatively low energy
($\sim 50$~keV; Maisack et al. 1993; Zdziarski et al. 1996), which may
reduce the importance of Compton heating.

NGC 3516 has been the subject of two major recent campaigns. The first
covered long time scales, where Maoz et al. (2000) showed a poor
zero-lag correlation between the X-ray and optical, with no clear
relationship at any lag.  There was possible evidence for the optical
leading the X-ray by $\sim 100$~d in the early part of the
observation, but this broke down in the later stages. The Maoz et
al. data could be explained by X-ray spectral variability, which is
clearly observed in NGC 3516 on long time scales. For example, Nandra
et al. (1999) noted a spectral index $\Gamma=1.5$ in a 1997 \asca\
observation, which can be compared to $\Gamma=1.9$ for observations
made in 1994-1995 (Nandra et al. 1997). The campaign of Edelson et
al. (2000) covered only a $\sim 3$~d period but was sampled very
intensively with \xte\ and {\it HST}.  The X--ray data showed
variations much larger in amplitude than those in the optical, which
can be accounted for by the light-travel smoothing we suggest for NGC
7469. On the other hand the optical did show small but significant
rapid variations, which showed no obvious relation to the 2-10 keV
X-ray variations. The amplitude of those variations was extremely
small, however, so they could merely represent variations (intrinsic
or via reprocessing) in the innermost part of the disk. In the case of
both long and short-term variations, an examination of the \xte\
spectral data for NGC 3516 could be most revealing.

In NGC 5548, Clavel et al. (1992) noted a correlation between the 2-10
keV X-rays and the UV, albeit with few points and poor
sampling. Marshall et al. (1997) showed that the EUV and UV variations
were also well correlated, consistent with photoelectric reprocessing
being a dominant effect. Completing the picture, Chiang et al. (2000)
show that the EUV variations are well correlated with those in the
2-10 keV band. These workers also show a correlation between $\Gamma$
and the 2-10 keV flux, and therefore presumably with the EUV/UV fluxes
as well. This behavior is entirely consistent with what we observe in
NGC 7469, and our model for that behavior.

\acknowledgements

We are grateful to the \xte\ GOF and PCA instrument teams for their
help and support in the data analysis.  KN is supported by NASA grant
NAG5-7067 to the Universities Space Research Association. BMP
acknowledges support by NASA LTSA grant NAG5-8397 to Ohio State
University.

\appendix
\section{The apparent correlation between $\Gamma$ and $A_{\rm ref}$}

As stated in the main text, parameters derived in a single spectral
fit can be correlated due to the presence of statistical errors.  This
is of particular concern in the case of the correlation between
$\Gamma$ and $A_{ref}$, which are highly correlated in the spectral
fits. To illustrate this further we plot in Fig.~\ref{fig:cont}
confidence contours of the two parameters for three of the spectra,
which are elongated in the direction of the correlation.

We have performed simulations to verify whether the observed
correlation between $\Gamma$ and $A_{\rm ref}$ is an artifact of the
analysis. First, we constructed 30 simulated spectra (i.e the number
of daily segments) with $\Gamma$ and $A_{\rm ref}$ covering the
observed ranges ($\Gamma=1.7-2.0$ and $A_{\rm ref}=0.002-0.010$). The
count rate was set at the mean value ($\sim 10$ ct s$^{-1}$) and the
exposure time at the mean for the daily segments. The 30 simulated
spectra were then fit with the same model as the real data and the
correlation coefficient calculated. We repeated this 100 times. The
mean linear correlation coefficient was found to be $r=0.12$ with a
standard deviation of 0.10. The maximum value obtained was 0.32.
These can be compared with the value derived from our data of
$r=0.75$. Very similar results were obtained for the rank correlation
coefficient. This shows that purely statistical effects are highly
unlikely to reproduce the observed correlation.

The spectral fits in Table~\ref{tab:var} are made under the assumption
that the observed power law slope is the same as that which
illuminates the material responsible for the Compton reflection. In
our case, as $\Gamma$ is variable, so is the shape of the reflection
component. If, in fact, the shape (and flux) of the Compton reflection
is constant, then our fitting could result in an apparently-variable
$A_{\rm ref}$, which is correlated with $\Gamma$. Physically, this
might be expected in a case where the reflecting material had a large
physical extend, in which case the medium would respond to the
``mean'' $\Gamma$ rather than the instantaneous value obtained in one
of our one day snapshots. Simulations confirm that this can produce
an apparent correlation between the parameters. If we make artificial
spectra with a constant reflection component with illuminating
$\Gamma=1.92$ (the mean value), then we find a clear correlation
between $\Gamma$ and $A_{\rm ref}$ when fitting with variable-$\Gamma$
model. Typical correlation coefficients are $r\sim 0.7$, similar
to that observed. 

An analogous situation is when there are systematic errors in the
background subtraction - particularly a systematic underestimate.
These can also cause an apparent correlation between the
parameters. We have simulated such an effect by making simulated
spectra with the parameters actually derived from the individual daily
segments, but before fitting, artificially modifying the background
spectrum so that it underestimates the flux at high energies.  In
particular we changed the PCA background spectrum such that the error
had a roughly power-law form, which rose from zero at low energies to
a maximum value of $\sim 3$~per cent at 20 keV.  This represents a
$\sim 30$~per cent error in the source-minus-background spectrum at
these energies. Repeating the 30-spectrum simulation 100 times, we
obtained correlation coefficients in the range 0.7-0.8, again
similar to the observed value of 0.75.

We conclude that the observed (and apparently highly significant)
correlation between $\Gamma$ and $A_{\rm ref}$ observed in our
spectral fits could easily be due either to a reflection component
with constant shape and flux, or to a systematic under-subtraction of
the background. 

\clearpage

\begin{deluxetable}{lllllll}

\tablecolumns{7}
\tablecaption{Spectral fits to daily segments
\label{tab:var}}

\tablehead{
\colhead{Day} & \colhead{Exposure} &
\colhead{$F_{\rm 2-10}$} & \colhead{$\Gamma$} &
\colhead{$F_{\rm K\alpha}$} & \colhead{$A_{\rm ref}$} &
\colhead{\chisq} \\
\colhead{TJD} & \colhead{ks} &
\colhead{$10^{-11}$} & \colhead{} & 
\colhead{$10^{-5}$}  & \colhead{$10^{-3}$} & 
\colhead{43 d.o.f.}
}
\startdata
245.104 & 18.8 & 3.43 &$1.973\pm 0.045$&$5.50\pm 0.96$&$6.4\pm 3.7$ & 47.9 \nl
246.177 & 18.4 & 3.45 &$1.979\pm 0.045$&$4.99\pm 0.97$&$5.3\pm 3.6$ & 48.5 \nl
247.247 & 16.1 & 2.89 &$1.929\pm 0.053$&$5.35\pm 1.00$&$2.3\pm 2.9$ & 44.1 \nl
248.316 & 15.5 & 3.72 &$1.919\pm 0.043$&$5.32\pm 1.07$&$4.2\pm 3.2$ & 65.2 \nl
249.386 & 14.8 & 3.81 &$1.928\pm 0.044$&$5.30\pm 1.10$&$4.0\pm 3.3$ & 64.6 \nl
250.456 & 13.1 & 3.92 &$1.902\pm 0.040$&$4.75\pm 1.17$&$1.9\pm 2.6$ & 39.8 \nl
251.579 & 16.5 & 3.40 &$1.832\pm 0.042$&$5.95\pm 1.02$&$2.8\pm 2.4$ & 70.4 \nl
252.631 & 16.1 & 3.17 &$1.860\pm 0.047$&$5.59\pm 1.02$&$4.1\pm 2.8$ & 43.4 \nl
253.676 & 15.0 & 2.55 &$1.707\pm 0.052$&$3.98\pm 1.00$&$1.2\pm 1.6$ & 53.1 \nl
254.714 & 14.5 & 3.17 &$1.874\pm 0.049$&$5.70\pm 1.07$&$2.5\pm 2.8$ & 46.7 \nl
255.777 & 12.5 & 2.81 &$1.909\pm 0.065$&$5.82\pm 1.13$&$5.8\pm 4.0$ & 63.3 \nl
256.811 & 15.2 & 2.76 &$1.853\pm 0.057$&$4.90\pm 1.02$&$5.1\pm 3.1$ & 49.7 \nl
257.840 & 13.9 & 2.75 &$1.882\pm 0.063$&$4.80\pm 1.07$&$6.5\pm 3.8$ & 43.0 \nl
258.865 & 15.8 & 2.55 &$1.930\pm 0.068$&$4.42\pm 0.99$&$6.9\pm 4.3$ & 50.8 \nl
259.914 & 16.6 & 2.87 &$1.980\pm 0.063$&$4.43\pm 0.99$&$10.2\pm 5.1$& 44.9 \nl
260.914 & 14.0 & 2.61 &$1.926\pm 0.069$&$4.83\pm 1.05$&$6.5\pm 4.3$ & 51.6 \nl
261.918 & 15.4 & 2.83 &$1.943\pm 0.060$&$5.40\pm 1.02$&$6.2\pm 4.0$ & 42.7 \nl
262.882 & 13.3 & 3.37 &$1.983\pm 0.056$&$4.99\pm 1.13$&$7.4\pm 4.8$ & 54.2 \nl
263.883 & 19.6 & 3.29 &$2.009\pm 0.050$&$4.47\pm 0.93$&$10.2\pm 4.6$& 49.7 \nl
264.917 & 15.6 & 3.53 &$1.959\pm 0.045$&$5.26\pm 1.05$&$3.7\pm 3.4$ & 41.2 \nl
265.953 & 18.2 & 3.91 &$1.957\pm 0.041$&$5.02\pm 1.00$&$7.8\pm 3.7$ & 68.4 \nl
267.019 & 18.2 & 4.04 &$1.899\pm 0.036$&$6.11\pm 1.01$&$3.6\pm 2.7$ & 61.2 \nl
268.084 & 18.4 & 3.93 &$1.939\pm 0.039$&$5.48\pm 1.00$&$6.6\pm 3.3$ & 60.2 \nl
269.155 & 19.4 & 3.86 &$1.867\pm 0.032$&$6.51\pm 0.97$&$1.2\pm 1.8$ & 66.7 \nl
270.222 & 16.5 & 3.33 &$1.847\pm 0.043$&$5.72\pm 1.01$&$2.9\pm 2.5$ & 44.9 \nl
271.282 & 17.9 & 3.41 &$1.821\pm 0.035$&$5.16\pm 0.98$&$1.0\pm 1.6$ & 62.8 \nl
272.361 & 16.5 & 2.72 &$1.774\pm 0.050$&$4.88\pm 0.97$&$2.2\pm 2.0$ & 57.6 \nl
273.417 & 20.1 & 2.66 &$1.872\pm 0.053$&$3.24\pm 0.87$&$5.3\pm 3.0$ & 59.9 \nl
274.488 & 19.4 & 3.22 &$1.902\pm 0.044$&$5.30\pm 0.93$&$5.3\pm 2.9$ & 47.5 \nl
275.547 & 21.4 & 3.22 &$2.014\pm 0.049$&$4.76\pm 0.89$&$10.3\pm 4.5$& 59.9 \nl

\tablecomments{$F_{2-10}$ is in units of erg cm$^{-2}$ s$^{-1}$;
$F_{\rm K\alpha}$ is in units of photon cm$^{-2}$ s$^{-1}$.
$A_{ref}$ is defined in the text.
}

\enddata
\end{deluxetable}

\begin{deluxetable}{llllllll}
\small

\tablecolumns{8}
\tablecaption{Correlations between the parameters \label{tab:cor}}

\tablehead{
\colhead{}              & \colhead{$F_{\rm 2-10}$} & 
\colhead{$F_{K\alpha}$} & \colhead{$\Gamma$} & 
\colhead{$A_{\rm ref}$} & \colhead{$F_{\rm UV}$} & 
\colhead{$Q_{0.1-100}$} & \colhead{$F_{UVextrap}$} 
}
\startdata
$F_{\rm 2-10}$&\nodata& {\it 0.54}  & 0.21  &-0.23  & 0.22  & {\it 0.73}  & 0.45  \nl
$F_{K\alpha}$ & {\it 0.52}  &\nodata& 0.01  &-0.34  & 0.02  & 0.25  & 0.04  \nl
$\Gamma$      & 0.13  &-0.20  &\nodata& {\it 0.75}  & {\it 0.81}  & {\it 0.81}  & {\it 0.92}  \nl
$A_{\rm ref}$ &-0.26  &-0.36  & {\it 0.75}  &\nodata& {\it 0.50}  & 0.41  & {\it 0.64}  \nl
$F_{\rm UV}$  & 0.26  &-0.05  & {\it 0.85}  & {\it 0.49}  &\nodata& {\it 0.71}  & {\it 0.80}  \nl
$Q_{0.1-100}$ & {\it 0.71}  & 0.18  & {\it 0.75}  & 0.37  & {\it 0.68}  &\nodata& {\it 0.94}  \nl 
$F_{UVextrap}$  & {\it 0.47}  &-0.02  & {\it 0.92}  & {\it 0.58}  & {\it 0.82}  & {\it 0.93}  &\nodata\nl
\tablecomments{
The upper right portion of the table shows the Pearson linear
correlation coefficient.  The lower left part shows the Spearman
(rank) correlation coefficients.  The correlations have 30 points,
except for those with $F_{\rm UV}$ which have 29.  Approximately,
values of $|r|>0.36,0.46,0.57$ are formally significant at
$>95,99,99.9$ per cent confidence. Those correlation coefficients
with formal chance probabilities $<1$~per cent are shown in italics,
although we note that these probabilities can be modified when
dealing with ``red noise'' light curves (see text).  
}
\enddata
\end{deluxetable}


\clearpage

\clearpage

\clearpage
\begin{figure}
\plotone{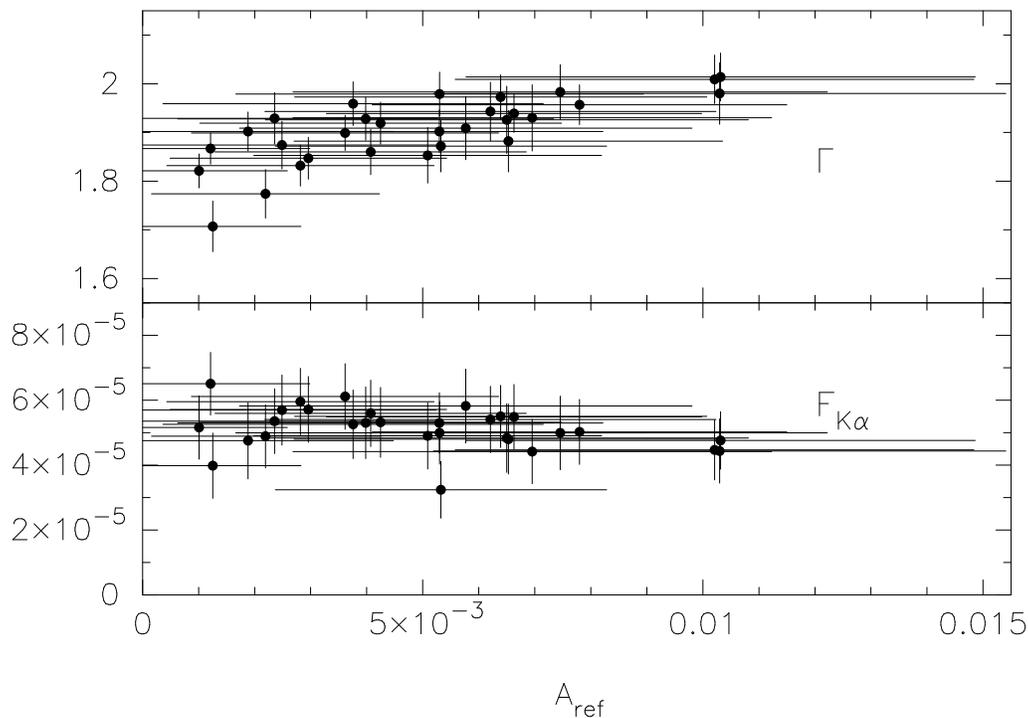}
\caption{$A_{\rm ref}$ plotted against $\Gamma$ (upper panel) and
$F_{\rm K\alpha}$ (lower panel). The former clearly shows a
correlation, although much of this could be statistical in nature, due
to the natural correlation between the parameters in the spectral fits
(Appendix A). The latter shows no significant relationship, even
though the components are thought to have a common origin as
reflection from the accretion disk. The apparent variability in
the reflection continuum may be due to a combination of statistical and
systematic effects, however, accounting for this decoupling.
\label{fig:aref}}
\end{figure}

\clearpage
\begin{figure}
\plotone{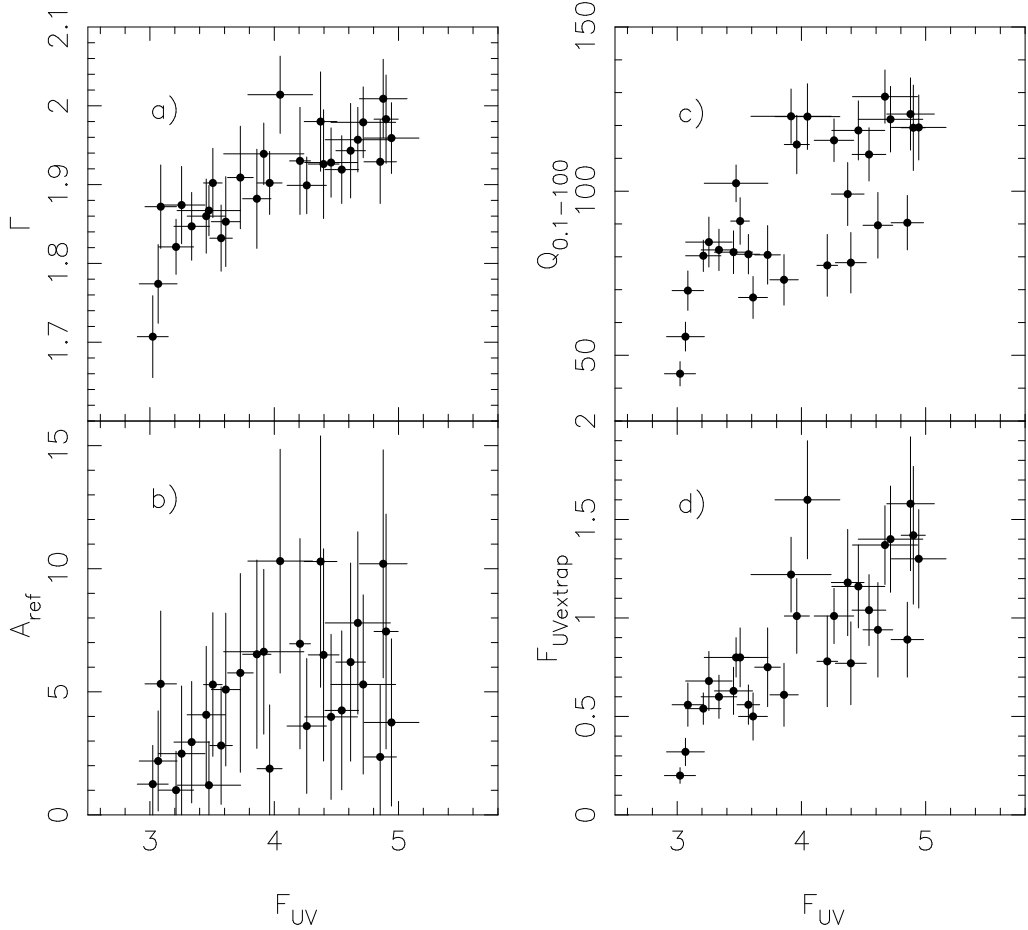}
\caption{Correlations between the UV continuum flux at 1315 \AA,
$F_{\rm UV}$ and a) the photon index $\Gamma$, b) the reflection
normalization, $A_{\rm ref}$ c) the broad-band X-ray {\it photon}
flux, $Q_{\rm 0.1-100}$ and d) the predicted UV continuum flux at 1315
A based on an extrapolation of the X-ray power law, $F_{\rm
UVextrap}$. The correlation with $A_{\rm ref}$ is rather marginal and
may be exaggerated by the effect that $A_{\rm ref}$ and $\Gamma$ are
correlated, which may be spurious (Appendix A). The other correlations
formally have very small chance probabilities ($>99.9$~per cent
confidence). The correlations of $F_{\rm UV}$ with $\Gamma$ and photon
flux strongly support the hypothesis that the X-ray flux arises from
Compton upscattering of UV seed photons (see text).  The final
correlation illustrates the more radical hypothesis that the variable
part of the UV flux may simply be an extrapolation of the X-ray power
law variabilitys.  Correlation coefficients are shown in
Table~\ref{tab:cor}.
\label{fig:ucor}}
\end{figure}

\clearpage
\begin{figure}
\plotone{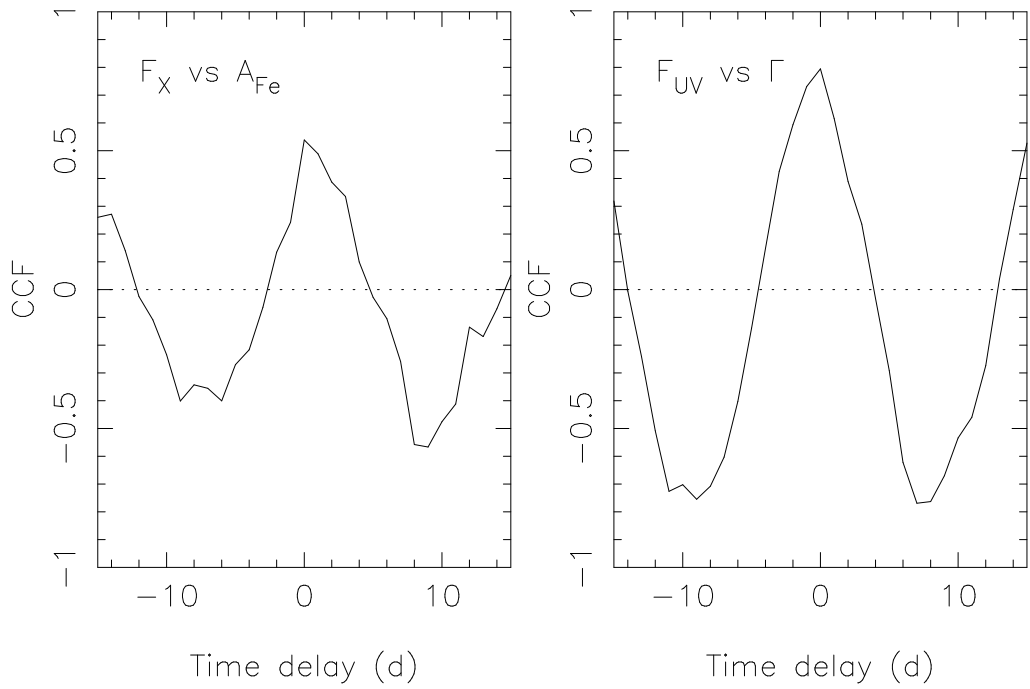}
\caption{Cross-correlation function for $F_{\rm K\alpha}$ versus
$F_{\rm 2-10}$ (left panel) and $\Gamma$ versus $F_{\rm UV}$ (right
panel).  Both show a strong peak at zero lag but the data quality and
sampling are not adequate to provide meaningful constraints or upper
limits on the time lag.
\label{fig:ccf}}
\end{figure}

\clearpage
\begin{figure}
\plotone{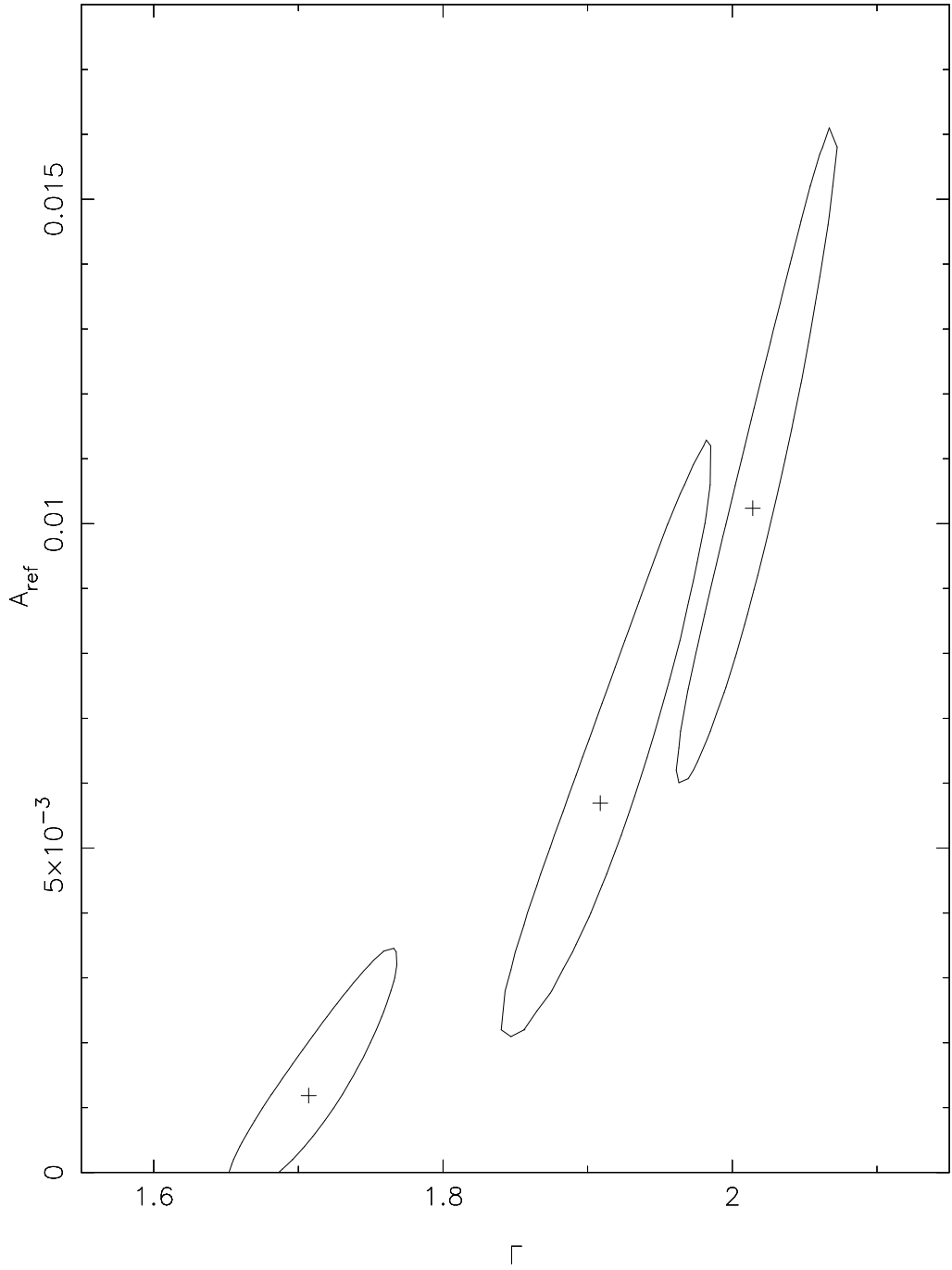}
\caption{90 per cent confidence contours for $\Gamma$ and $A_{\rm ref}$ 
for the spectra of segments 9, 11 and 30, which cover the range 
of these parameters. The two parameters are highly correlated
in the individual fits and this, combined with possible systematic
effects, can cause the apparent correlation observed between the
parameters seen in Fig.~\ref{fig:aref} (Appendix A)
\label{fig:cont}}
\end{figure}

\end{document}